\documentstyle[12pt,epsf]{article}
 
\begin{document}

\baselineskip 6mm
\renewcommand{\thefootnote}{\fnsymbol{footnote}}


\newcommand{\nc}{\newcommand}
\newcommand{\rnc}{\renewcommand}


\rnc{\baselinestretch}{1.24}    
\setlength{\jot}{6pt}       
\rnc{\arraystretch}{1.24}   

\makeatletter
\rnc{\theequation}{\thesection.\arabic{equation}}
\@addtoreset{equation}{section}
\makeatother



\nc{\be}{\begin{equation}}

\nc{\ee}{\end{equation}}

\nc{\bea}{\begin{eqnarray}}

\nc{\eea}{\end{eqnarray}}

\nc{\xx}{\nonumber\\}

\nc{\ct}{\cite}

\nc{\la}{\label}

\nc{\eq}[1]{(\ref{#1})}

\nc{\newcaption}[1]{\centerline{\parbox{6in}{\caption{#1}}}}

\nc{\fig}[3]{

\begin{figure}
\centerline{\epsfxsize=#1\epsfbox{#2.eps}}
\newcaption{#3. \label{#2}}
\end{figure}
}


\def\CA{{\cal A}}
\def\CC{{\cal C}}
\def\CD{{\cal D}}
\def\CE{{\cal E}}
\def\CF{{\cal F}}
\def\CG{{\cal G}}
\def\CH{{\cal H}}
\def\CK{{\cal K}}
\def\CL{{\cal L}}
\def\CM{{\cal M}}
\def\CN{{\cal N}}
\def\CO{{\cal O}}
\def\CP{{\cal P}}
\def\CR{{\cal R}}
\def\CS{{\cal S}}
\def\CU{{\cal U}}
\def\CW{{\cal W}}
\def\CY{{\cal Y}}


\def\IR{{\hbox{{\rm I}\kern-.2em\hbox{\rm R}}}}
\def\IB{{\hbox{{\rm I}\kern-.2em\hbox{\rm B}}}}
\def\IN{{\hbox{{\rm I}\kern-.2em\hbox{\rm N}}}}
\def\IC{\,\,{\hbox{{\rm I}\kern-.59em\hbox{\bf C}}}}
\def\IZ{{\hbox{{\rm Z}\kern-.4em\hbox{\rm Z}}}}
\def\IP{{\hbox{{\rm I}\kern-.2em\hbox{\rm P}}}}
\def\IH{{\hbox{{\rm I}\kern-.2em\hbox{\rm H}}}}
\def\ID{{\hbox{{\rm I}\kern-.2em\hbox{\rm D}}}}


\def\a{\alpha}
\def\b{\beta}
\def\ga{\gamma}
\def\d{\delta}
\def\ep{\epsilon}
\def\ph{\phi}
\def\k{\kappa}
\def\l{\lambda}
\def\m{\mu}
\def\n{\nu}
\def\th{\theta}
\def\rh{\rho}
\def\s{\sigma}
\def\t{\tau}
\def\w{\omega}
\def\G{\Gamma}


\def\half{\frac{1}{2}}
\def\dint#1#2{\int\limits_{#1}^{#2}}
\def\goto{\rightarrow}
\def\para{\parallel}
\def\brac#1{\langle #1 \rangle}
\def\grad{\nabla}
\def\curl{\nabla\times}
\def\div{\nabla\cdot}
\def\p{\partial}
\def\e{\epsilon_0}


\def\Tr{{\rm Tr}\,}
\def\det{{\rm det}}


\def\vare{\varepsilon}
\def\barz{\bar{z}}
\def\barw{\bar{w}}


\def\ad{\dot{a}}
\def\bd{\dot{b}}
\def\cd{\dot{c}}
\def\dd{\dot{d}}
\def\so{SO(4)}
\def\sop{SO(4)^\prime}
\def\bc{{\bf C}}
\def\bfz{{\bf Z}}
\def\bz{\bar{z}}

\begin{titlepage}


\hfill\parbox{3.7cm} {HU-EP-05/65 \\
{\tt hep-th/0510249}}

\vspace{15mm}

\begin{center}
{\Large \bf ALE spaces from noncommutative U(1) instantons via exact Seiberg-Witten map}

\vspace{10mm}
Mario Salizzoni\footnote{sali@physik.hu-berlin.de},
Alessandro Torrielli\footnote{torriell@physik.hu-berlin.de}
and Hyun Seok Yang\footnote{hsyang@physik.hu-berlin.de}
\\[10mm]

{\sl Institut f\"ur Physik, Humboldt Universit\"at zu Berlin \\
Newtonstra\ss e 15, D-12489 Berlin, Germany}

\end{center}

\thispagestyle{empty}

\vskip1cm


\centerline{\bf ABSTRACT}
\vskip 4mm
\noindent

The exact Seiberg-Witten (SW) map of a noncommutative (NC) gauge theory gives
the commutative equivalent as an ordinary gauge theory
coupled to a field dependent effective metric. We study instanton
solutions of this commutative equivalent whose self-duality
equation turns out to be the exact SW map of
NC instantons. We derive general differential
equations governing $U(1)$ instantons and we explicitly get an exact
solution corresponding to the single NC instanton.
Remarkably the effective metric induced by the single $U(1)$
instanton is related to the Eguchi-Hanson metric$-$the
simplest gravitational instanton. Surprisingly the instanton
number is not quantized but depends on an integration constant.
Our result confirms the expected non-perturbative breakdown of the
SW map. However, the breakdown of the map arises in
a consistent way: The instanton number plays the 
role of a parameter giving rise to a one-parameter family of 
Eguchi-Hanson metrics.
\\



\vspace{1cm}

\today

\end{titlepage}

\renewcommand{\thefootnote}{\arabic{footnote}}
\setcounter{footnote}{0}

\section{Introduction}

Noncommutative (NC) spaces can be obtained by quantizing a given
space with a symplectic structure $\theta_{\mu\nu}$:
\be \la{nc-space}
[x^\mu, x^\nu]_\star = i \theta_{\mu\nu},
\ee
where star-product is defined by
\begin{equation}\label{star-product}
(f \star g)(x) = \left.\exp\left(\frac{i}{2}\theta^{\mu\nu}
\partial_{\mu}^{x}\partial_{\nu}^{y}\right)f(x)g(y)\right|_{x=y}.
\end{equation}
Also field theories can be formulated on the NC space. NC field
theory means that fields are defined as functions over the NC
space, whose products are defined by the NC star-product \eq{star-product}. 
At the algebraic level, the fields become operators acting
on the Hilbert space as a representation space of Eq.\eq{nc-space}.

On such space, the exponential $e^{i k \cdot x}$ acts as a translation
operator, i.e., 
\begin{equation}
e^{i k \cdot x} \star f(x) \star e^{- i k \cdot x} = f(x+ k \cdot \theta),
\end{equation}
showing that a translation in a NC direction is equivalent to a gauge
transformation up to a global symmetry transformation. This property is shared 
with another physical theory, General Relativity, where translations are
also equivalent to gauge transformations. It is then natural to 
investigate the structure of general relativity inherent 
in the far simpler NC field theories. 
Our current work supports that NC gauge theories are really good toy models of
general relativity, as asserted in \ct{quote}.

To derive the action that governs NC gauge theories, we recall that they
naturally arise as a decoupling limit of the open string dynamics on D-branes
in the Neveu-Schwarz $B$ field background. In the limit of slowly varying
fields, the open string effective action on a D-brane is given by the
Dirac-Born-Infeld (DBI) action \ct{dbi}.
Seiberg and Witten, however, showed \ct{sw}
that an explicit form of the effective action depends on the
regularization scheme of the two dimensional field theory defined by
the worldsheet action. That is, depending on the regularization
scheme or path integral prescription for the open string ending on
a D-brane, one can have two descriptions: commutative and noncommutative 
descriptions. Since these two descriptions arise from the same
open string theory and since the physics should not depend on the regularization
scheme, it was argued in \ct{sw} that the two descriptions 
should be equivalent and thus there must be a spacetime field redefinition between
ordinary and NC gauge fields, the so called Seiberg-Witten (SW)
map. In this sense NC gauge theories have a dual description
through the SW map in terms of ordinary gauge theories on
commutative spacetime. To understand the dual description exactly,
it is important to know the exact SW map.

If one uses the commutative description via the SW map, however, the
connection between translation and gauge transformation is lost.
A global translation on commutative fields can no longer be
written as a gauge transformation. So one may wonder how the properties
related to gravity in NC gauge theories show up in the commutative
description via the SW map. It turns out \ct{rivelles,yang} that, 
when the commutative description is employed, an ``effective metric'' induced
by gauge fields directly emerges. Of course, this property suggests 
a possible gravitational interpretation. 
We will show that there exists a rigorous way of establishing this
connection. More precisely, we will see that the effective metric generated by
the single NC $U(1)$ instanton is related to the Eguchi-Hanson (EH) 
metric$-$the simplest gravitational instanton \ct{eh}.

The paper is organized as follows. In Section 2, we briefly summarize the
exact SW map obtained in \ct{yang,ban-yang}.
In Section 3, we derive the Bogomoln'yi bound for the 
commutative action obtained from the NC theory via the exact SW map.
The resulting self-duality equation turns out to be the exact SW 
map of NC instantons.
We derive the general differential equations governing $U(1)$ instantons and
explicitly get an exact solution corresponding to a single NC 
instanton. We show that the instanton number is surprisingly not quantized
but depends on an integration constant. This result shows a non-perturbative
breakdown of the SW map whose possibility was already anticipated
in \ct{sw} by Seiberg and Witten and in \ct{harvey}, more rigorously, by Harvey.
In Section 4, we observe the remarkable fact that the single NC $U(1)$
instanton is mapped to the EH space.
Furthermore, the breakdown of the SW map arises in a
consistent way: The instanton number plays the
role of a parameter giving rise to a one-parameter family of EH metrics.
Since our effective metric is K\"ahler, 
we find the K\"ahler potential of the effective metric whose physical meaning 
from the gauge theory side is not yet obvious. 
Finally, in Section 5, we briefly summarize our results and discuss related open issues.

\section{Exact Seiberg-Witten map of noncommutative gauge theory}

The action for NC electrodynamics in flat Euclidean ${\bf R}^4$ is given by
\begin{equation}\label{nced}
\widehat{S}_{\mathrm{NC}} = \frac{1}{4}\int\! d^4 x \,\widehat{F}_{\mu\nu} \star \widehat{F}^{\mu\nu},
\end{equation}
where noncommutative fields are defined by
\be \la{ncf}
\widehat{F}_{\mu\nu}=\partial_{\mu}\widehat{A}_{\nu}-\partial_{\nu}\widehat{A}_{\mu}-
i \,[\widehat{A}_{\mu}, \widehat{A}_{\nu}]_\star.
\ee
For the reason mentioned in the Introduction, there should be a
commutative deformed electrodynamics equivalent to Eq.\eq{nced}. 
It was shown in \ct{yang,ban-yang}, for slowly varying fields on a single
D-brane, that the dual description of the NC DBI action 
through the exact SW map is simply given by the ordinary DBI action
expressed in terms of open string variables:
\begin{eqnarray}\label{sw-equiv}
 &&  \int d^{p+1} x \sqrt{\det(G + \kappa (\widehat{F}+ \Phi))} \nonumber \\
 && \hspace{2cm} = \int d^{p+1} x \sqrt{\det{(1+ F \theta})}
 \sqrt{\det{(G + \kappa (\Phi + {\bf F}))}} +   {\cal O}(\sqrt{\kappa} \partial F),
\end{eqnarray}
where
\begin{equation}\label{def-fatf}
    {\bf F}_{\mu\nu} (x) \equiv \Biggl(\frac{1}{1 + F\theta} F\Biggr)_{\mu\nu} (x)
\end{equation}
and
\begin{equation}\label{cf}
F_{\mu\nu} (x) = \p_\mu A_\nu(x) -  \p_\nu A_\mu(x).
\end{equation}
The commutative action in Eq.\eq{sw-equiv} is exactly the same as
the DBI action obtained from the worldsheet sigma model using 
$\zeta$-function regularization \ct{andreev}.
For a closed string background characterized by $B_{\mu\nu}, g_{\mu\nu}$ and $g_s$,
we have a continuum of descriptions labelled by a choice of
$\Phi$. In the following, we are interested in $\Phi =0$, the familiar NC
description, the open string metric $G_{\mu\nu} =\delta_{\mu\nu}$ and $p=3$.

In the zero slope limit $\kappa \equiv 2 \pi \alpha^\prime \to 0$, 
one can expand both sides of
Eq.\eq{sw-equiv} in terms of powers of $\kappa$ and produce infinitely many
identities related to each other by the exact SW map. At ${\cal O}(\kappa^2)$,
we get the exact commutative nonlinear electrodynamics equivalent to
Eq.\eq{nced}
\begin{equation}\label{ced-sw}
S_{\mathrm{C}} = \frac{1}{4} \int d^4 x \sqrt{\det{{\rm g}}} \;
{\rm g}^{\mu \alpha} {\rm g}^{\beta\nu} F_{\mu\nu}
F_{\alpha\beta},
\end{equation}
where we introduced an ``effective metric" induced
by the dynamical gauge fields such that
\begin{equation}\label{effective-metric}
    {\rm g}_{\mu\nu} = \delta_{\mu\nu} + (F\theta)_{\mu\nu},
    \qquad  ({\rm g}^{-1})^{\mu\nu} \equiv {\rm g}^{\mu\nu} =
    \Biggl(\frac{1}{1 + F\theta}\Biggr)^{\mu\nu}.
\end{equation}
Note that the effective metric \eq{effective-metric} is in general not symmetric. 
The action \eq{ced-sw} is very interesting in the sense
that the NC electrodynamics after the exact SW map can be regarded
as the ordinary electrodynamics coupled to the ``effective metric"
$\mathrm{g}_{\mu\nu}$ \ct{rivelles}.
It should be remarked, however, that the effective metric in the action \eq{ced-sw}
cannot be interpreted just as a fixed background since it depends
on the dynamical gauge field.
It is easy to derive the exact equation of motion from the action \eq{ced-sw}
\bea \la{eom-exact}
&& \p_\mu \Biggl[ \sqrt{-{\rm g}} \biggl\{ (\theta {\rm g}^{-1})^{\mu\alpha}
\Tr ({\rm g}^{-1} F {\rm g}^{-1} F) - 2 \Bigl ( (\theta{\rm
g}^{-1}F {\rm g}^{-1} F {\rm g}^{-1})^{\mu\alpha} - (\theta{\rm
g}^{-1}F {\rm g}^{-1} F {\rm g}^{-1})^{\alpha\mu}\Bigr) \xx
&& \hspace{1.7cm} + 2 \Bigl( ({\rm g}^{-1}F {\rm g}^{-1})^{\mu\alpha}
- ({\rm g}^{-1}F {\rm g}^{-1})^{\alpha\mu} \Bigr) \biggr\}
\Biggr] = 0.
\eea
It was checked in \ct{yang} that the nonlinear action Eq.\eq{ced-sw} is consistent
with the results in \ct{berrino} where it was proved that the terms of order
$n$ in $\theta$ in the action via the SW map form a
homogeneous polynomial of degree $n+2$ in $F$ 
and explicitly presented the deformed action up to order $\theta^2$.

The commutative action \eq{ced-sw} can actually be also derived
from the NC action \eq{nced} using the exact SW map in
\ct{ban-yang,jur-sch,liu} (see \ct{liu,oka-oog} for the exact inverse SW map):
\bea \label{eswmap}
    &&  \widehat{F}_{\mu\nu}(x) = \Biggl(\frac{1}{1 + F\theta} F
    \Biggr)_{\mu\nu}(X), \\
    \la{measure-sw}
    && d^{4} x = d^{4} X \sqrt{\det(1+ F \theta)}(X),
\eea
where
\begin{equation}\label{X}
    X^\mu(x) \equiv x^\mu + \theta^{\mu\nu}
    \widehat{A}_\nu(x).
\end{equation}
This is consistent with the above mentioned result by \ct{berrino}. 
Here we used different coordinates, $X^\mu$ and $x^\mu$, for commutative and
NC descriptions, respectively. However we will often use the symbol $x$ for
both descriptions whenever the distinction is not necessary.

The exact SW map between topological invariants was also found in
\ct{ban-yang}. For example, using the SW maps \eq{eswmap} and
\eq{measure-sw} and the identity
\be \la{top-id}
\int d^4 x \sqrt{\det{\rm g}}
\Bigl({\rm g}^{-1} F \Bigr) \wedge \Bigl({\rm g}^{-1} F \Bigr)
=  \int d^4 x F  \wedge F,
\ee
where the wedge notation has been used
\begin{equation}
    F \wedge F \equiv
    \varepsilon^{\mu\nu \lambda\rho} F_{\mu\nu} F_{\lambda\rho}, \nonumber
\end{equation}
one can get the exact SW map between instanton numbers
\be \la{instanton-sw}
\int d^4 x (\widehat{F} \wedge \widehat{F})(x) = \int d^4 X (F \wedge F)(X).
\ee
The identity \eq{top-id} will play a crucial role to derive the Bogomoln'yi bound of the action
\eq{ced-sw}. The proof of the identity \eq{top-id} is simple if one notices that the quantity
$\Bigl({\rm g}^{-1} F \Bigr)_{\mu\nu}$ is anti-symmetric: 
\be \la{derivation-top-id}
\sqrt{\det{\rm g}}
\Bigl({\rm g}^{-1} F \Bigr) \wedge \Bigl({\rm g}^{-1} F\Bigr)
= 8 \sqrt{\det{\rm g}} \, {\mathrm Pf}({\rm g}^{-1} F)
= 8 \sqrt{\det{\rm g}}\, \sqrt{\det({\rm g}^{-1} F)}
= 8 {\mathrm Pf}F= F  \wedge F.
\ee
In next section, we will see that the identity \eq{instanton-sw} is
interestingly broken, suggesting a non-perturbative breakdown
of the SW map. We will further comment on this property 
of the map in Section 5.

\section{Exact Seiberg-Witten map of noncommutative instantons}

${\bf R}^4$ is the simplest hyper-K\"ahler manifold, viewed as
the quaternions $\IH \simeq \bf{C}^2$. We introduce the quaternions $\IH$ defined by
\bea \label{quaternion1}
 && {\bf x}=x_\mu \sigma^\mu = \left(%
\begin{array}{cc}
  x_4 + i x_3 & x_2 + i x_1 \\
  -x_2 + i x_1 & x_4 - i x_3 \\
\end{array}%
\right)
= \left(%
\begin{array}{cc}
  z_2 & z_1 \\
  -\bar{z}_1 & \bar{z}_2 \\
\end{array}%
\right) \\
\la{quaternion2}
&& {\bar{\bf x}}=x_\mu {\bar\sigma}^\mu
= \left(%
\begin{array}{cc}
  x_4 - i x_3 & -x_2 - i x_1 \\
  x_2 - i x_1 & x_4 + i x_3 \\
\end{array}%
\right)
= \left(%
\begin{array}{cc}
 \bar{z}_2 & - z_1 \\
  \bar{z}_1 & z_2 \\
\end{array}%
\right)
\eea
where $\sigma^\mu=(i \tau^a, 1)$ and ${\bar \sigma}^\mu=(-i
\tau^a, 1)=-\sigma^2 \sigma^{\mu T} \sigma^2$.
The quaternion matrices $\sigma^\mu$
and ${\bar \sigma}^\mu$ have the basic properties
\bea \la{sigma}
&& \sigma^\mu {\bar \sigma}^\nu=\delta^{\mu\nu}+i\sigma^{\mu\nu},
\qquad \sigma^{\mu\nu}=\eta^a_{\mu\nu} \tau^a=*\sigma^{\mu\nu},\xx
&& {\bar \sigma}^\mu \sigma^\nu=\delta^{\mu\nu}+i{\bar
\sigma}^{\mu\nu}, \qquad {\bar \sigma}^{\mu\nu}={\bar
\eta}^a_{\mu\nu} \tau^a= -*{\bar \sigma}^{\mu\nu},
\eea
where the $4 \times 4$ matrices $\eta^a_{\mu\nu}$ and ${\bar \eta}^a_{\mu\nu}$ are
't Hooft symbols defined by
\bea \la{tHooft-symbol}
&& {\bar \eta}^a_{ij} = {\eta}^a_{ij} = {\varepsilon}_{aij}, \qquad i,j \in
\{1,2,3\}, \nonumber\\
&& {\bar \eta}^a_{4i} = {\eta}^a_{i4} = \delta_{ai}.
\eea
We list some identities of the 't Hooft tensors that will be useful for later calculations:
\bea
\la{self-eta}
&& \eta^{(\pm)a}_{\mu\nu} = \pm \half \varepsilon_{\mu\nu\lambda\kappa}
\eta^{(\pm)a}_{\lambda\kappa}, \\
\la{proj-eta}
&& \eta^{(\pm)a}_{\mu\nu}\eta^{(\pm)a}_{\lambda\kappa}
= \delta_{\mu\lambda}\delta_{\nu\kappa}
-\delta_{\mu\kappa}\delta_{\nu\lambda} \pm
\vare_{\mu\nu\lambda\kappa}, \\
\la{eta-etabar}
&& \eta^{(\pm)a}_{\mu\nu} \eta^{(\mp)b}_{\mu\nu}=0, \\
\la{eta^2}
&& \eta^{(\pm)a}_{\lambda\mu}\eta^{(\pm)b}_{\lambda\nu}
=\delta_{ab}\delta_{\mu\nu}+\vare_{abc}\eta^{(\pm)c}_{\mu\nu},
\eea
where $\eta^{(+)a}_{\mu\nu} = \eta^a_{\mu\nu}$ and
$\eta^{(-)a}_{\mu\nu} = {\bar \eta}^a_{\mu\nu}$.

The Euclidean Lorentz group $O(4)$ is isomorphic to $SU(2)_L \times SU(2)_R$,
where two $SU(2)$ subgroups correspond to the left-handed and
right-handed chiral rotations. The $O(4)$ group acts on the
quaternion ${\bf x}$ as
\be \la{o4}
{\bf x} \; \longrightarrow \; g_L {\bf x} g_R.
\ee
The self-dual (SD) and anti-self-dual (ASD) two-forms are
basically given by the triple of K\"ahler 2-forms over $\IH$
\bea \la{sd-2form}
&& \omega^a_{SD} = - \frac{i}{4}{\mathrm tr}_2 (\tau^a d{\bf x} \wedge
d{\bar{\bf x}}), \\
&& \omega^a_{ASD} = - \frac{i}{4}{\mathrm tr}_2
(\tau^a d{\bar{\bf x}} \wedge d{\bf x}),
\eea
where ${\mathrm tr}_2$ denotes the trace over quaternionic
indices. Note that only one part of Lorentz symmetry acts on the
sphere of complex structures of the hyper-K\"ahler manifold $\IH \simeq
\bf{C}^2$:
\bea \la{complex-sphere}
&& \omega^a_{SD} \rightarrow - \frac{i}{4}{\mathrm tr}_2
(g_L^\dagger \tau^a g_L d{\bf x} \wedge d{\bar{\bf x}}), \\
&& \omega^a_{ASD} \rightarrow - \frac{i}{4}{\mathrm tr}_2
(g_R \tau^a g_R^\dagger d{\bar{\bf x}} \wedge d{\bf x}).
\eea

By the noncommutativity \eq{nc-space}, the original Lorentz symmetry
is broken down to its subgroup. For the SD and ASD $\theta_{\mu\nu}$,
the original Lorentz symmetry $O(4) \cong SU(2)_L \times SU(2)_R$ is
broken down to the subgroup
\begin{equation}\label{lsb}
  SU(2)_L \times SU(2)_R \rightarrow
\left \{ \begin{array}{l} SU(2)_R \times U(1)_L,
\qquad \mbox{SD},\\
  SU(2)_L \times U(1)_R, \qquad \mbox{ASD}.
\end{array} \right.
\end{equation}
Now we will restrict to the self-dual NC ${\bf R}^4$, with the canonical form 
$\theta_{\mu\nu}= \frac{\theta}{2} \eta^3_{\mu\nu}$. In this case
the moduli space of the Lorentz symmetry breaking \eq{lsb} is
parameterized by $SU(2)_L/U(1)_L \cong {\bf S}^2$, which can be
regarded as the Hopf map $\pi: {\bf S}^3 \rightarrow {\bf
S}^2$ \ct{yang02}. Using quaternions, the standard Hopf map can be
represented as
\be \la{hopf-map}
T^a = - \frac{1}{4}{\mathrm tr}_2
(\tau^3 {\bf x} \tau^a \bar{\bf x}).
\ee
In terms of ${\bf C}^2$ and ${\bf R}^4$ variables, they are
explicitly given by
\bea \la{t123}
&& T^1 = - \half (z_1 \bar{z}_2 + \bar{z}_1 z_2) = -(x_1x_3 + x_2
x_4), \xx
&& T^2 = - \frac{i}{2} (z_1 \bar{z}_2 - \bar{z}_1 z_2) = x_1x_4 - x_2
x_3, \\
&& T^3 = \half (z_1 \bar{z}_1 - z_2 \bar{z}_2) = \half (x_1^2 + x_2^2 -
x_3^2-x_4^2) \nonumber
\eea
and
\be \la{t^2}
\sum_{a=1}^3 T_a^2 = \frac{1}{4}r^4
\ee
with $r^2 = z_1 \bar{z}_1 + z_2 \bar{z}_2$. Under the Lorentz
transformation \eq{o4}, they transform as
\be \la{04-t}
T^a  \rightarrow - \frac{1}{4}{\mathrm tr}_2
(g_L^\dagger \tau^3 g_L {\bf x} (g_R \tau^a g_R^\dagger) \bar{\bf x}).
\ee

Since instanton solutions are the field configuration satisfying Bogomoln'yi
bound, we also expect that the commutative instantons 
we want to find from the action \eq{ced-sw}
similarly satisfy the corresponding Bogomoln'yi bound of Eq.\eq{ced-sw}.
Thus our problem is how to find the self-duality equation of Eq.\eq{ced-sw}
by applying the Bogomoln'yi trick. An essential hint comes 
from the fact that the left-hand side of Eq.\eq{top-id} is a topological term
defining instanton number on commutative ${\bf R}^4$. Guided
by this fact, we rewrite the action \eq{ced-sw} in the form
\be \label{bogo}
S_{\mathrm{C}} = \frac{1}{8}
\int d^4 x \sqrt{\det{{\rm g}}} \Bigl({\bf F}_{\mu\nu} \mp \half
\varepsilon_{\mu\nu\alpha\beta}{\bf F}_{\alpha\beta} \Bigr)^2
\pm \frac{1}{8} \int d^4 x \sqrt{\det{\rm g}}
\Bigl({\rm g}^{-1} F \Bigr) \wedge \Bigl({\rm g}^{-1} F\Bigr).
\ee
Note that the first term is positive definite since $d^4 x \sqrt{\det{{\rm
g}}}$ is anyway the volume form of the NC coordinates
(see Eq.\eq{measure-sw}) and so positive definite,
while the second term is topological because of Eq.\eq{top-id}
and thus does not affect the equations of motion. So we propose
the self-duality equation for the action $S_{\mathrm{C}}$ to be 
\be \la{c-self-dual}
{\bf F}_{\mu\nu} (X) = \pm \half
\varepsilon_{\mu\nu\alpha\beta}{\bf F}_{\alpha\beta} (X).
\ee
Note that the above equation is directly obtained by applying the exact SW map \eq{eswmap}
to the NC self-duality equation, i.e.,
\be \la{nc-self-dual}
{\widehat F}_{\mu\nu} (x) = \pm \half
\varepsilon_{\mu\nu\alpha\beta}{\widehat F}_{\alpha\beta} (x).
\ee

Using Eq.\eq{top-id}, one can check that the field configuration that 
satisfies the self-duality equation \eq{c-self-dual} also satisfies the
equation of motion \eq{eom-exact}: Take a variation with respect
to the gauge field $A_\mu$ on both sides of Eq.\eq{top-id} and then the right-hand
side identically vanishes. If the relation \eq{c-self-dual} is
substituted to Eq.\eq{eom-exact}, the result is equal to the
variation of the left-hand side of Eq.\eq{top-id} with respect
to $A_\mu$ and thus vanishes.

Since we want to consider a commutative equivalent of the
Nekrasov-Schwarz instanton \ct{nek-sch}, we will
consider the anti-self-dual case in Eq.\eq{c-self-dual}.
To solve the equation \eq{c-self-dual}, we will take
the following general strategy.

(I) Take a general ansatz with the ASD two-form basis
$\omega^a_{ASD}$ as follows
\be \la{bff-ansatz}
{\bf F} \equiv  \half {\bf F}_{\mu\nu} dx^\mu \wedge dx^\nu =
f^a(x)\omega^a_{ASD},
\ee
where $f^a$'s are arbitrary functions. Then the equation
\eq{c-self-dual} is automatically satisfied.

(II) Solve the field strength $F_{\mu\nu}$ in terms of ${\bf
F}_{\mu\nu}$:
\begin{equation}\label{f-fatf}
    F_{\mu\nu} (x) = \Biggl(\frac{1}{1-{\bf F}\theta}{\bf F}\Biggr)_{\mu\nu}(x).
\end{equation}
Then impose the Bianchi identity for $F_{\mu\nu}$,
\be \la{bianchi}
\varepsilon_{\mu\nu\rho\sigma}\partial_{\nu} F_{\rho\sigma} = 0,
\ee
since the field strength $F_{\mu\nu}$
is given by a (locally) exact two-form, i.e., $F=dA$. In the end
we will get general differential equations governing $U(1)$
instantons.

Substituting the ansatz \eq{bff-ansatz}, ${\bf F}_{\mu\nu}(x) =
f^a(x)\bar{\eta}^a_{\mu\nu}$, into Eq.\eq{f-fatf}, we get
\be \la{field-f}
F_{\mu\nu} = \frac{1}{1-\phi}f^a \bar{\eta}^a_{\mu\nu} -
\frac{2\phi}{\theta(1-\phi)}\eta^3_{\mu\nu},
\ee
where
\be \la{Phi}
\phi \equiv \frac{\theta^2}{4}\sum_{a=1}^3 f^a(x)f^a (x).
\ee
Using the result \eq{field-f} and Eq.\eq{self-eta}, the Bianchi identity \eq{bianchi}
is reduced to the following differential equations
\be \la{diff-eq}
f^a \bar{\eta}^a_{\mu\nu} \partial_\nu \phi + (1-\phi)
\bar{\eta}^a_{\mu\nu} \partial_\nu f^a + \theta^{-1} \eta^3_{\mu\nu}
\partial_\nu \phi^2 - \theta^{-1} \eta^3_{\mu\nu} \partial_\nu (1-\phi)^2 =0.
\ee
From Eq.\eq{field-f}, we obtain
\be \la{sw-instanton}
F_{\mu\nu}^+ \equiv \half(F_{\mu\nu}+ \half \varepsilon_{\mu\nu\rho\sigma}
F_{\rho\sigma}) = \frac{1}{4}(F\widetilde{F}) \theta_{\mu\nu}^+
\ee
since
\be \la{fdualf}
F\widetilde{F} \equiv \half
\varepsilon^{\mu\nu\rho\sigma}F_{\mu\nu}F_{\rho\sigma}=
- \frac{16 \phi}{\theta^2(1-\phi)}.
\ee
Note that Eq.\eq{sw-instanton} is precisely the instanton equation,
Eq.(4.54) in \ct{sw}, used and explicitly solved there for single instanton case. 
See also \ct{3ms-tera}. Thus the instanton solution (4.60)
in \ct{sw} was interestingly the exact solution although they got Eq.\eq{sw-instanton}
perturbatively.

We will solve the differential equation \eq{diff-eq} for the single instanton case.
We will set $\theta = 1$ from now on, but it can be easily recovered by a 
simple dimensional analysis by recalling that $\theta$ carries the dimension
of $(length)^2$. Since our instanton equation \eq{c-self-dual} (which is more fundamental than
Eq.\eq{sw-instanton}) is obtained by the SW map from
Eq.\eq{nc-self-dual}, we take an ansatz of the same form as the NC
instanton \ct{nek-sch,yang99}
\be \la{ansatz1}
f^a(x) = f(r) T^a.
\ee
It is straightforward to derive an ordinary differential equation
for the function $f(r)$ from Eq.\eq{diff-eq}:
\be \la{abel-ode}
r(r^2 f + 4) \frac{df}{dr} - 2(r^2 f - 12)f=0,
\ee
where we assumed $(r^2 f + 4)\neq 0$ which turns
out to be true. To get the result \eq{abel-ode}, the
following relations might be useful
\be \la{2relations}
\bar{\eta}^a_{\mu\nu} \partial_\nu T^a = 3 \eta^3_{\mu\nu} x^\nu,
\qquad \bar{\eta}^a_{\mu\nu} x^\nu T^a = \frac{r^2}{2} \eta^3_{\mu\nu}
x^\nu.
\ee

Eq.\eq{abel-ode} is the well-known Abel's ordinary differential equation of the
second kind. One may test asymptotic behaviors of the solution by
assuming them as $f(r) = c r^{-n}$. The result is that $f(r)=
4/r^2$ when $r
\rightarrow 0$ while $f(r) \sim c/r^6$ or $cr^2$ when $r
\rightarrow \infty$. But we will require that the solution has to
rapidly decay at $r \rightarrow \infty$ and thus we need
$f(r) \sim c/r^6$ at $r \rightarrow \infty$. The exact solutions
are given by
\be \la{abel-sol}
f(r) = \frac{4}{r^2} \frac{\sqrt{1 + \frac{t^4}{r^4}} \mp 1}
{\sqrt{1 + \frac{t^4}{r^4}} \pm 1},
\ee
where $t^4$ is an integration constant. Only the upper sign
solution satisfies the correct asymptotic behavior.

The corresponding gauge field $A_\mu(x)$ can be found by taking the most general
$SU(2)_R \times U(1)_L$ invariant ansatz \ct{sw}
\be \la{ansatz-a}
A_\mu(x) = \eta^3_{\mu\nu} x^\nu h(r).
\ee
The expression about the field strength $F_{\mu\nu}$ can be found
in Eq.(4.56) in \ct{sw}:
\bea \la{4.56-sw}
&& F_{12} = -2h -(x_1^2 + x_2^2) \frac{h^\prime}{r},
\qquad \quad \;\; F_{34} = -2h -(x_3^2 + x_4^2) \frac{h^\prime}{r}, \xx
&& F_{13} = F_{24} = (x_1 x_4 - x_2 x_3) \frac{h^\prime}{r},
\qquad F_{23} = - F_{14}= (x_2x_4 + x_1 x_3) \frac{h^\prime}{r}.
\eea
By comparing Eq.\eq{4.56-sw} with Eq.\eq{field-f} with the ansatz
\eq{ansatz1}, we can find the following relations
\be \la{h-f}
\frac{1}{1-\frac{r^4}{16} f^2} f = - \frac{h^\prime}{r}, \qquad
\frac{r^4}{8(1-\frac{r^4}{16} f^2)} f^2 = 2h + \frac{r}{2}
h^\prime.
\ee
From these relations, we get
\be \la{hfromf}
h(r) = \frac{r^2 f}{4 - r^2 f} = - \half + \half \sqrt{1 + \frac{t^4}{r^4}}
\ee
and, substituting Eq.\eq{hfromf} into Eq.\eq{h-f}, we get the 
differential equation \eq{abel-ode} again. 
Conversely, the equation \eq{abel-ode} is equivalent to
\be \la{diff-h}
r \frac{d}{dr}(h^2 + h) = -4 (h^2 + h).
\ee

Although the field strength $F_{\mu\nu}$ contains a (mild)
singularity  at $r=0$, ${\bf F}_{\mu\nu}$ in Eq.\eq{def-fatf} is
completely non-singular. Since the mild singularity in $\sqrt{\det(1+F\theta)} \sim
t^2/4r^2$ for $r \rightarrow 0$ can be safely compensated by the volume term
$d^4 x$, the action \eq{ced-sw} does not contain any harmful
singularities. This behavior near $r=0$ is exactly what is needed
to give the solution a finite and nonzero instanton number. Let us
calculate the instanton number:
\be \la{instanton-number}
I \equiv \frac{1}{32\pi^2} \int d^4 x F\widetilde{F} =
-\frac{1}{16}\int_{0}^{\infty} dr \frac{r^7f^2}{1-\frac{1}{16}r^4f^2}
= - \frac{t^4}{16}.
\ee
Surprisingly the instanton number depends on the integration
constant $t^4$. In a sense this fact was already shown in \ct{sw}
since Seiberg and Witten solved exactly the same instanton
equation \eq{sw-instanton} as ours. Since our approach used 
the exact SW map, i.e., included all corrections from $\theta$-dependent
terms, this result shows a non-perturbative breakdown
of the SW map. However, this possibility of the map was
already anticipated in \ct{sw,seiberg} 
and, more rigorously, in \ct{harvey}, where it was 
pointed out that commutative and NC gauge fields have different
topology. Indeed there is no topological reason that commutative
$U(1)$ instantons have a quantized topological charge.

The behavior in the commutative limit, i.e., $\theta \to 0$, can be clarified
by recovering the dimensionful parameter $\theta$.
In particular, one has to set $t^2 = \theta \widetilde{t}^2$ with a dimensionless constant
$\widetilde{t}$. In the limit of small $\theta$ and fixed $\widetilde{t}$, 
the solution \eq{abel-sol} goes to zero, while the quantity $I$ in
Eq.\eq{instanton-number} remains constant. 
In this way, one does not recover the strictly
commutative case. We notice, however, that fixing the quantity
$\theta \widetilde{t}^4$ and sending $\theta$ to zero gives
$f(r) \sim \theta \widetilde{t}^4/r^6$ (an additional $\theta^{-1}$ appears in front of
Eq.\eq{abel-sol} after dimensional analysis).
The Nekrasov-Schwarz instanton \ct{nek-sch} 
also exhibits the same behavior in the $\theta \to 0$ limit. 
This behavior, $f(r) \sim C/r^6$, is also proper of the solution 
in the strictly commutative case, with a fixed dimensionful constant $C$. 
Taking the commutative limit, we thus get $ t^2 \to 0$ while 
$\widetilde{t}^4 \to \infty$. We recover in this way the result 
for the solution of the strictly linear equation $F^+_{\mu\nu} = 0$ 
with the same symmetry \ct{sw} where $I \propto  \widetilde{t}^4 \to \infty$. 

\section{ALE spaces from $U(1)$ instantons}

In the Introduction, we speculated about the possibility 
that NC gauge fields can play the role of gravity. 
In this section, we will try to give a precise
mathematical and physical connection between $U(1)$ instantons on
${\bf R}^4$ with flat metric and gravitational instantons,
which are hyper-K\"ahler four-manifolds arising in general relativity.

Let us rewrite the ``effective metric" in Eq.\eq{effective-metric} 
in the form \footnote{Since the effective metric 
${\rm g}_{\mu\nu} = (1 - {\bf F}\theta)^{-1}_{\mu\nu}$, 
the metric determined by Eq.\eq{c-self-dual} is symmetric if 
$\varepsilon^{\mu\nu\rho\sigma} {\bf F}_{\mu\nu} \theta_{\rho\sigma}=0$.}  
\be \la{new-metric}
{\rm g}_{\mu\nu} = \half (\delta_{\mu\nu} + \widetilde{{\rm g}}_{\mu\nu}).
\ee 
It is straightforward to get the metric $\widetilde{{\rm g}}_{\mu\nu}$ 
using Eq.\eq{4.56-sw} with the solution
\eq{hfromf}:
\be \la{eh-metric}
\widetilde{{\rm g}}_{\mu\nu} = \left(%
\begin{array}{cccc}
  G_1 & 0 & G_4 & G_3 \\
  0 & G_1 & -G_3 & G_4 \\
  G_4 & -G_3 & G_2 & 0 \\
  G_3 & G_4 & 0 & G_2 \\
\end{array}%
\right),
\ee
where
\bea \la{comp-eh}
&& G_1 = G - H (x_1^2 + x_2^2),
\qquad \;\; G_{2} = G - H (x_3^2 + x_4^2), \xx
&& G_3 = - H(x_1 x_4 - x_2 x_3),
\qquad G_{4} = - H (x_2x_4 + x_1 x_3) \nonumber
\eea
and
$$
G = \frac{\sqrt{r^4 + t^4}}{r^2}, \qquad
H = -\frac{G^\prime}{2r}= \frac{t^4}{r^4\sqrt{r^4 + t^4}}.
$$
First we note that $\widetilde{{\rm g}}_{\mu\nu} = \delta_{\mu\nu} + {\cal O}(r^{-4})$ at infinity,
which is a common property of a particular family of
hyper-K\"ahler manifold, the so-called ALE spaces \ct{kronh,joyce}. 
Indeed this metric is precisely the EH metric, the simplest ALE
space,\footnote{It may be worthwhile to point out
that the isometry group of the instanton and the EH space 
is also coincident with $SU(2)_R \times U(1)_L$.} in the form that can be
found in \ct{pablo}. We thus constructed the gravitational instanton from $U(1)$ gauge
fields in the {\it flat} spacetime. 
The integration constant $t^4$ plays the role of a parameter giving rise to a
one-parameter family of EH metrics. This metric
becomes flat in the case of $t=0$ (except at the origin). Thus the family of the
EH space is parameterized by the instanton number !

The instanton number is now endowed with a meaning by the K\"ahler geometry. 
So the $U(1)$ instantons are consistently connected to the hyper-K\"ahler
geometries. In this way, commutative $U(1)$ instantons
behave as a source that generates K\"ahler geometries. 
If one insists on the (anti-)self-dual
configurations of the action \eq{ced-sw}, it thus looks like a theory of
K\"ahler geometry rather than a theory of gauge fields.

Since our effective metric, $ds^2 = \widetilde{{\mathrm g}}_{i\bar{j}} dz_i
d\bar{z}_j,\; i,j=1,2$, is hyper-K\"ahler, we can introduce a K\"ahler two-form
$\Omega$ and a K\"ahler potential $K$ defined by
\be \la{kaehler-potential}
\Omega = \frac{i}{2} \widetilde{{\mathrm g}}_{i\bar{j}} dz_i \wedge d\bar{z}_j
= \frac{i}{2} \partial \bar{\partial} K
\ee
where the exterior derivative is defined by
\be
d= dx^\mu \frac{\partial}{\partial x^\mu} = dz^i \partial_i
+ d\bar{z}^i \bar{\partial}_i = \partial +
\bar{\partial}.
\nonumber
\ee
Using the definition \eq{kaehler-potential}, one can easily check
that the K\"ahler potential $K$ \ct{lebrun,joyce,pablo} for the EH metric
\eq{eh-metric} is given by
\be \la{pot-k}
K= \sqrt{r^4 + t^4} + t^2 \log \frac{r^2}{\sqrt{r^4 + t^4} + t^2}.
\ee
It is very remarkable for the $U(1)$ instanton to reproduce precisely
the K\"ahler potential for the EH metric.
However, it is not yet obvious what is the physical meaning of the K\"ahler potential
from the gauge theory side although it is an important ingredient
in the K\"ahler geometry. We leave this interpretation for future work.

\section{Discussion}

We studied the commutative instantons related to NC instantons
by the exact SW map.\footnote{The SW map of NC instantons was previously studied
in \ct{hash-oog} for localized intantons generated by shift operators
and in \ct{kra-shig} for the Nekrasov-Schwarz instantons.}
We found self-duality equations, from which we got general differential
equations governing $U(1)$ instantons. We observed that our self-duality
equation is equivalent to the instanton equation in \ct{sw}.
We also found that the instanton number is no longer quantized.
This result suggests a non-perturbative breakdown of the SW map.
Let us further discuss this last property.

The SW map is a map between gauge orbit spaces of commutative and
NC gauge fields. However, it was pointed out \ct{sw,seiberg} that
the change of variables from $\widehat{A}_\mu$ to $A_\mu$ or vice
versa has only a finite radius of convergence. Thus the SW map
cannot completely encode the topology of gauge fields. Indeed
it was shown \ct{harvey} that the gauge orbit spaces for commutative and
noncommutative gauge theories are different. In particular, the topology of NC $U(1)$
gauge fields is nontrivial while the commutative one is trivial. Our
result confirms these differences. The other example of this
breakdown is the level quantization of NC Chern-Simons
theory for $U(1)$ gauge group \ct{nc-cs}. (For the exact SW map of the
Chern-Simons theory, see \ct{grandi}.) However this seems to be
interrelated to the instanton case due to the following relation \ct{ban-yang}
\begin{equation}\label{chern}
    \int d^{4} x \widehat{F} \wedge \widehat{F} = \int d^{4} x \; d \widehat{\Omega}_{\mathrm{CS}}
\end{equation}
with
\begin{equation}\label{chern-simons}
\widehat{\Omega}_{\mathrm{CS}} = \int^1_0 dt \widehat{A} \wedge \widehat{F}_t
\end{equation}
where $\widehat{F}_t = t d\widehat{A} -i t^2 \widehat{A} \star \widehat{A}$.
We derived the identity \eq{instanton-sw} from the exact SW map. But we
observed that the identity \eq{instanton-sw} is broken down: The
left-hand side carries a nontrivial topology while the right-hand
side carries trivial one. So, if one introduces a three-manifold ${\cal M}_3$ such
that $\partial {\cal M}_4 = {\cal M}_3$, one gets a different character of level
quantization for the commutative and NC Chern-Simons theories.

We showed that our effective metric induced by the commutative $U(1)$ instanton
is interestingly related to a particular family of hyper-K\"ahler manifold,
the so-called ALE spaces, at least for the simplest instanton.
We thus constructed the gravitational instanton from $U(1)$ gauge fields 
in {\it flat} spacetime.
Note that there is a general construction of all ALE manifolds by Kronheimer
\ct{kronh}. In this construction ALE spaces are explicitly obtained as
hyper-K\"ahler quotients of flat Euclidean spaces \ct{hklr} and emerge as minimal
resolutions of ${\bf C}^2/\Gamma$, where $\Gamma$ is a discrete subgroup of
$SU(2)$. For ALE spaces of A-type, corresponding to $\Gamma = \IZ_N$, the
metric is known to be diffeomorphic to the Gibbons-Hawking multi-center metric
\ct{gh-metric}. (The two-center Gibbons-Hawking metric is the EH
metric and the hyper-K\"ahler quotient construction of the EH
metric is given in the Appendix in \ct{ale}.) So it is a very interesting
problem to explore whether more general ALE spaces can be obtained by $U(1)$
instantons in the way we examined in this paper. Note that the EH
metric was obtained by the very special ansatz \eq{ansatz1} which is the Hopf
map with unit Hopf invariant. One may try an ansatz described by a Hopf
map with higher Hopf invariants. Conversely, one may try to find an instanton
solution corresponding to, for example, the multi-center Gibbons-Hawking
metric.

Another interesting problem is how to embed the commutative $U(1)$ instantons
described by Eq.\eq{c-self-dual} into the ADHM construction.
At first sight, this seems not possible since the instanton number is not
quantized. However, if the instanton solution of Eq.\eq{c-self-dual} is
generally related to the ALE spaces, one may rather extract the instanton
solutions from the Kronheimer's hyper-K\"ahler quotient construction 
of the ALE spaces \ct{kronh}. In this construction, the instanton number
appears as a deformation parameter in the moment map $\mu_{\bf R}$;
$\mu_{\bf R}= t^2$ \ct{ale}. Setting the deformation parameter $ t^2 = 0$, we get the
singular orbifold ${\bf C}^2/\Gamma$, consistent with the metric \eq{eh-metric}. 
The commutative limit discussed at the end of Section 3 thus corresponds to
the singlular limit of ALE spaces, as naturally expected result.

ALE spaces carry two topological invariants: the Euler
characteristic and the Hirzebruch signature \ct{egh-pr}.
A natural question arising from our work is what is the meaning of these
topological invariants from the gauge theory point of view. Since they should be
represented by higher derivative terms of $U(1)$ field strength $F_{\mu\nu}$,
they are very exotic objects in the gauge theory picture. We hope to address some of
the problems raised here in the near future.

\newpage

\section*{Acknowledgments}

We would like to thank Harald Dorn for very useful discussions at various
stages of this project and for reading the manuscript. 
DFG supported M.S. under the project SA 1356/1 
and A.T. within the ``Schwerpunktprogramm Stringtheorie 1096''. 
The work of H.S.Y. is supported by the Alexander von Humboldt
Foundation and he also thanks the staff of the Institut f\"ur Physik,
Humboldt Universit\"at zu Berlin for the cordial hospitality.


\nc{\npb}[3]{Nucl. Phys. {\bf B#1} (#2) #3}

\nc{\plb}[3]{Phys. Lett. {\bf B#1} (#2) #3}

\nc{\prl}[3]{Phys. Rev. Lett. {\bf #1} (#2) #3}

\nc{\prd}[3]{Phys. Rev. {\bf D#1} (#2) #3}

\nc{\ap}[3]{Ann. Phys. {\bf #1} (#2) #3}

\nc{\prep}[3]{Phys. Rep. {\bf #1} (#2) #3}

\nc{\epj}[3]{Eur. Phys. J. {\bf #1} (#2) #3}

\nc{\ptp}[3]{Prog. Theor. Phys. {\bf #1} (#2) #3}

\nc{\rmp}[3]{Rev. Mod. Phys. {\bf #1} (#2) #3}

\nc{\cmp}[3]{Comm. Math. Phys. {\bf #1} (#2) #3}

\nc{\mpl}[3]{Mod. Phys. Lett. {\bf #1} (#2) #3}

\nc{\cqg}[3]{Class. Quant. Grav. {\bf #1} (#2) #3}

\nc{\jhep}[3]{J. High Energy Phys. {\bf #1} (#2) #3}

\nc{\atmp}[3]{Adv. Theor. Math. Phys. {\bf #1} (#2) #3}

\nc{\hepth}[1]{{\tt hep-th/{#1}}}


\end{document}